\documentclass[review]{elsarticle}

\usepackage{lineno,hyperref}
\modulolinenumbers[5]

\journal{Journal of \LaTeX\ Templates}

\usepackage[plain]{fancyref}
\usepackage[dvipsnames]{xcolor}
\usepackage{algorithm2e}
\usepackage{caption}
\usepackage{subcaption}
\usepackage{flafter}
\usepackage{csquotes}
\usepackage{amsmath,amsfonts,bm}
\newcommand{\toolname}{TBSSvis}
\usepackage{pgf}					
\usepackage{tikz}					
\usepackage{varwidth}
\usetikzlibrary{calc}
\usetikzlibrary{%
  arrows,%
  shapes,
  chains,%
  matrix,%
  positioning,
  scopes,%
  decorations.pathmorphing,
  shadows,%
  calc
}\tikzset{
     textWcolor/.style={
        		circle,
            fill=white, 
            draw=black, 
            line width=1pt,
            font=\sffamily\scriptsize,
            opacity=0.7,
            inner sep=0pt,
            minimum size=10pt
    },
    textWborder/.style={
        		circle,
            fill=none,
            draw=black, 
            line width=1pt,
            font=\normalfont\bfseries\small,
            opacity=0.7,
            inner sep=0pt,
            minimum size=12pt
    },
    textOnly/.style={
        		rectangle,
        		fill=none, 
            font=\normalfont\scriptsize,
            inner sep=0pt,
            minimum size=10pt
    },
    cycleLabels/.style={
        		rectangle,
            fill=none,
            draw=none,
            font=\normalfont\small,
            inner sep=0pt,
            anchor=west
    },
    mathnode/.style={
        		rectangle,
            fill=none,
            draw=none,
            font=\normalfont\normalsize,
            inner sep=0pt,
            anchor=west
    },
    sfnode/.style={
        		rectangle,
            fill=none,
            draw=none,
            font=\sffamily\small,
            inner sep=0pt,
            anchor=west
    },
    image overlay index/.style={
        		circle,
            fill=col2,
            draw=black!40,
            line width=0.5pt,
            font=\normalfont\scriptsize,
            opacity=0.7,
            inner sep=0pt,
            minimum size=10pt
    }
}

\begin{document}

\begin{frontmatter}

\title{TBSSvis: Visual Analytics for Temporal Blind Source Separation}

\author[tuwieninf]{Nikolaus Piccolotto\corref{correspondingauthor}}
\cortext[correspondingauthor]{Corresponding author}
\author[tuwieninf]{Markus Bögl}
\author[erstegroup]{Theresia Gschwandtner}
\author[tuwienstat]{Christoph Muehlmann}
\author[yliopisto]{Klaus Nordhausen}
\author[tuwienstat]{Peter Filzmoser}
\author[tuwieninf]{Silvia Miksch}

\address[tuwieninf]{TU Wien, Institute of Visual Computing \& Human-Centered Technology, Favoritenstrasse 9–11, A-1040 Vienna}
\address[erstegroup]{Erste Group Bank AG, Am Belvedere 1, A-1100 Vienna}
\address[tuwienstat]{TU Wien, Institute of Statistics and Mathematical Methods in Economics, Wiedner Hauptstrasse 8–10, A-1040 Vienna}
\address[yliopisto]{University of Jyväskylä, Department of Mathematics and Statistics, FI-40014 Jyväskylä}

\begin{abstract}
Temporal Blind Source Separation (TBSS) is used to obtain the true underlying processes from noisy temporal multivariate data, such as electrocardiograms. TBSS has similarities to Principal Component Analysis (PCA) as it separates the input data into univariate components and is applicable to suitable datasets from various domains, such as medicine, finance, or civil engineering. Despite TBSS's broad applicability, the involved tasks are not well supported in current tools, which offer only text-based interactions and single static images. Analysts are limited in analyzing and comparing obtained results, which consist of diverse data such as matrices and sets of time series. Additionally, parameter settings have a big impact on separation performance, but as a consequence of improper tooling, analysts currently do not consider the whole parameter space. We propose to solve these problems by applying visual analytics (VA) principles. Our primary contribution is a design study for TBSS, which so far has not been explored by the visualization community. We developed a task abstraction and visualization design in a user-centered design process. Task-specific assembling of well-established visualization techniques and algorithms to gain insights in the TBSS processes is our secondary contribution. We present TBSSvis, an interactive web-based VA prototype, which we evaluated extensively in two interviews with five TBSS experts. Feedback and observations from these interviews show that TBSSvis supports the actual workflow and combination of interactive visualizations that facilitate the tasks involved in analyzing TBSS results. 
\end{abstract}


\end{frontmatter}


\section{Introduction}
\label{sec:introduction}

Multivariate measurements of a phenomenon are common in many domains. Medical doctors place electrodes on a patient's body to analyze processes such as brain activity, eye movements, or heart rhythm. Civil engineers measure vibrations on different parts of a structure, such as a bridge, to detect possible faults.
Financial managers invest money in stocks, which are in a way sensors of economic processes, to gain wealth. Common to all these examples is the time-oriented data and the assumption that data from different sensors is in some way correlated and/or influenced by noise. However, analysts are usually only interested in the ``true'' underlying processes.

To obtain these processes, analysts turn to Blind Source Separation (BSS). BSS comprises established methods for signal separation that were applied, among others, in the mentioned domains of medicine \cite{comon2010, delathauwer2000, vanthanh2017}, civil engineering \cite{amezquita-sanchez2016} and finance \cite{oja2000}. Temporal Blind Source Separation (TBSS) refers to a subset of BSS methods that specifically account for temporal correlation. TBSS is similar to Principal Component Analysis (PCA) in the sense that i) TBSS methods work on any multivariate dataset with quantitative variables, ii) they work on measured data only (hence ``blind'') and iii) separate it into a linear combination of uncorrelated components, like PCA. Unlike PCA, TBSS accounts for temporal correlation and often requires complex tuning parameters. As both TBSS and PCA can be considered forms of dimension reduction, analysts use TBSS and PCA for similar reasons, like data analysis or modeling/prediction. 

During these activities, it is at some point necessary to inspect components visually. Like with PCA, components are hidden until the separation algorithm is executed, but TBSS's complex parameter space severely complicates the issue: It is known that parameter settings greatly influence the result, but not in which way a change in parameters translates to change in components. Experts regard automated analysis by extensive sampling \cite{sedlmair2014} not a feasible option and there is little guidance from the literature, which parameters to pick. Because a ground truth is rarely available, TBSS analysis is inherently open-ended and exploratory as there are no known insights to confirm. The workflow of TBSS analysts can broadly be described as i) pick a parameter setting, ii) see if obtained components are useful or interesting and if not, go to i).

Some challenges make TBSS difficult to use in practice. Despite the important role of visualization in their workflow, the current tool used by the analysts does not support them well in this regard. Analysts need to manually program static visualizations, which requires time they could otherwise spend on data analysis.
Another challenge is the amount of components. Each parametrization on a $p$-variate dataset yields a set of $p$ components that need inspection and comparison to previous sets. Analysts are, for example, interested in commonly found components, but very quickly confronted with hundreds of components to consider. This is a common task in ensemble visualization \cite{wang2019}, but made more difficult by components appearing in sets instead of one by one.
Also, when comparing multiple results, analysts will eventually find competing options for their final choice. As there is usually no ground truth available to compare the result to, analysts need detailed ways to compare individual results to make an informed decision.

Visual analytics (VA~\cite{thomas2005}) as defined by Keim et al. \cite{keim2008} ``combines automated analysis techniques with interactive visualizations for an effective understanding, reasoning and decision making on the basis of very large and complex data sets''. Considering the strong focus of BSS analysis on visual inspection on multiple levels of detail, in combination with mentioned challenges, we propose applying VA principles to overcome these. We designed \toolname{} according to Munzner's Nested Model \cite{munzner2009} for the TBSS method ``generalized Second Order Blind Identification'' (gSOBI) \cite{miettinen2020}. We chose gSOBI because it is recent and well suited to real-world datasets due to its flexibility (see \Fref{sec:related-work--tbss}).  The source code of \toolname{} is available at \url{https://github.com/npiccolotto/tbss-vis}.


Our primary research contribution is a design study \cite{sedlmair2016} for TBSS, which improves the visualization community's knowledge about an area that it did not explore so far. Specifically we provide:

\begin{itemize}
  \item A task abstraction for TBSS which we obtained through a user-centered visualization design process with TBSS experts (\Fref{sec:task-abstraction}).
  \item A VA design for gSOBI, a TBSS method, that supports the abstracted tasks by combining visualizations, interactions, and guidance methods (\Fref{sec:app}).
  \item Confirmation of the effectiveness of our design in two interviews with five TBSS experts (\Fref{sec:evaluation}).
\end{itemize}

As part of this design study we put well-established visualization techniques together to support the identified tasks. These, together with a set-aware clustering scheme (\Fref{sec:app--compare-comonents}), are our secondary contribution. They include a multivariate autocorrelation function plot and the application of a slope graph to sets of time series.

\section{Related Work}

In the following we elaborate on different approaches to visualize and compare time series, ensembles, and models.





\subsection{Time Series Visualization}

Temporal data is ubiquitous in many domains such as finance, health, or biology, and has been visualized for centuries since the first line graph was introduced by Playfair \cite{tufte2001}. Various other visual encodings have been proposed afterwards, such as tile maps, sparklines, or horizon graphs \cite{aigner2011}. They use different visual variables \cite{mackinlay1986} such as position, color, or slope, and therefore exhibit different perceptual properties, which makes them suitable for different analysis tasks. E.g., Gogolou et al. \cite{gogolou2019} investigated the relation between different time series visualization idioms and perceived similarity. They recommend to use horizon graphs when local variations in temporal position or speed is important, while others (line graph, color band) are better suited for notions of similarity where amplitude is less important. 


Many time series can be visualized by juxtaposition, as is the case in LiveRAC \cite{mclachlan2008}. Various system measures (columns) are displayed per machine (rows) in a space-filling table design, using semantic zooming to change the level of detail between color bars, sparklines, and labeled line graphs. When not using all available space, one could use small multiples \cite{tufte2001} in different arrangements. For instance, Stitz et al. \cite{stitz2016} arrange small multiples of stocks by price and price change in a user-selected time frame. Liu et al. \cite{liu2018a}, on the other hand, lay them out with a modified Multidimensional Scaling (MDS) algorithm such that similar items are near each other.



Superimposed encodings trade decreased usage of display space for legibility, as they do not scale well after a couple of variables due to occlusion. An example besides the well known superimposed line graph is the braided graph \cite{javed2010}, which superimposes multiple area-based marks. Because of the varying data dimensionality in TBSS, superimposition is generally not a promising strategy. 

To keep features of long time series visible, designers often turn to focus-and-context techniques such as lenses \cite{tominski2017}. In the simplest case, a lens mainly enlarges an area of interest, such as in SignalLens \cite{kincaid2010}. But more complex interactions are possible, such as in ChronoLenses \cite{zhao2011}, where users can combine and stack multiple lenses.
An alternative to interaction is to reduce the visualized data either in a data-driven \cite{shurkhovetskyy2018} or visualization-driven way, e.g., by line simplification \cite{rosen2020}. However, if expected features in the data are not known in advance, as is the case in TBSS, one risks that important features are removed.

\subsection{Ensemble Visualization}

The goal in ensemble visualization is to make sense of a set of similar complex data items, such as trajectories, often produced by a simulation with perturbed parameter settings. Component sets obtained from different TBSS parametrizations constitute such an ensemble. Ensemble visualization has its origin in meteorology \cite{potter2009}, but since expanded to more domains \cite{wang2019}. 
Analytic tasks for ensemble data \cite{wang2019} indicate popular strategies, such as comparing members or grouping them by similarity, to support the stated goal. Existing works \cite{hao2016, ferstl2017} often use popular clustering techniques (with domain-specific distance functions) to support the latter task. This is not straightforward in TBSS as one has to take care to not mix members of different sets into the same cluster. We discuss our approach in \Fref{sec:app--compare-comonents}.

Time is a common part of ensemble data, but not a requirement \cite{matkovic2009, piringer2012, matkovic2018, xu2019}. One possible case is when ensemble members are univariate time series, such as for Köthur et al. \cite{kothur2015}, who encoded the correlation between members in a heatmap to support comparison of two ensembles. More commonly, other data types have an associated time dimension such as multivariate data \cite{obermaier2016}, particle data \cite{hao2016}, network security data \cite{hao2015}, or spatial data \cite{buchmuller2019}.

\subsection{VA for Construction and Comparison of Models}

Sedlmair et al. \cite{sedlmair2014} devise a framework for visual parameter analysis of models, but when applied, analysts still have to make themselves familiar with the solution space. As there is often no a-priori goal, model construction can be characterized as exploratory analysis.

VA has supported the construction of different kinds of models such as linear regression \cite{zhao2014}, logistic regression \cite{dingen2019}, time series \cite{bogl2013}, dimension reduction \cite{anand2012},  or classification \cite{choo2010} and also related tasks like variable selection \cite{krause2014}.

Reducing available models to a set of candidates and choosing the final model are tasks that require comparisons of models. The final choice depends on many factors. Recently, VA tools have been developed to compare machine learning models, to ensure their predictions are fair and free of bias \cite{zhang2019a, wexler2020}. Comparison tools for demand forecasting \cite{sun2020}, decision tree \cite{muhlbacher2018}, and regression \cite{muhlbacher2013} models exist too. 

\section{Temporal Blind Source Separation}
\label{sec:related-work--tbss}

The statistical analysis of multiple measurements taken at different times is a challenging task. Often, such multivariate time series are analyzed by transforming the data in certain simple ways to uncover latent processes which generated the data. Probably the most used method for such a task is the classical PCA, which uses linear transformations of the data that result in components which have highest variance and are uncorrelated. Uncorrelated implies that the covariance between the found linear combinations is zero. The linear combinations are given by diagonalizing the covariance matrix. Furthermore, as the nature of the transformation is linear, interpretations of the results can be carried out by the simple and well studied loadings-scores scheme. However, PCA might not be the best choice when the data at hand shows dependencies in time, as the main source of information is in that case not covariance, but rather serial dependence. Serial dependence is characterized by autocovariance, i.e., covariance between measurements separated in time by a given lag. In analogy to PCA it would be desirable to find linear combinations of the multivariate time series data which are not only uncorrelated marginally (zero covariance between variables at each time step), but also uncorrelated in time (zero autocovariance between variables for any lag). TBSS is a field of multivariate statistics that studies methods delivering the former desired properties. Generally, BSS is a well established model-based framework. It assumes that the observed data are a linear mixture of latent components, which are considered usually easier to model and/or more meaningful for interpretation than multivariate models. The goal of BSS is to recover these components based on the observed data alone. BSS is formulated and used for many types of data, as outlined in recent reviews \cite{comon2010,NordhausenOja2018,YanMatilainenTaskinenNordhausen2021,NordhausenRuzgazen2022}. In the following we outline the concept of TBSS.

The model of TBSS considered here is
$
	\boldsymbol x_t = \boldsymbol A \boldsymbol c_t,
$
where $\boldsymbol x_t$ denotes the observed $p$-variate time series, $\boldsymbol A$ is the full-rank $p \times p$ mixing matrix and $\boldsymbol{c}_t= (c_{1,t},\ldots,c_{p,t})^\top$ is the set of $p$ latent components, which should be estimated. Thus the goal is to find a  $p \times p$ unmixing matrix $\boldsymbol W = (\boldsymbol w_1, \ldots, \boldsymbol w_p)^\top$, such that  $\boldsymbol c_t = \boldsymbol W \boldsymbol x_t$ up to sign and order of the components in $\boldsymbol c_t$. To facilitate the recovery, the assumption is made that the components in $\boldsymbol c_t$ have $\text{Cov}(\boldsymbol c_t)= \boldsymbol I_p$ and are uncorrelated (or independent) with mutually distinct serial dependence. This means, for example, that all cross-moment matrices, such as autocovariance matrices, of $\boldsymbol c_t$ are diagonal matrices.

A very first approach for TBSS is denoted as Second-Order Blind Identification (SOBI) algorithm \cite{belouchrani1997,MIETTINEN2014214,miettinen2016}. It finds the linear combinations of the data which make autocovariance matrices for several lags as diagonal as possible. Hence, found components are uncorrelated marginally and uncorrelated in time. It is well known in the statistical analysis of time series data, that time series emerging from different scientific fields have different key characteristics. For example, financial time series are not well characterized by autocovariance matrices, but instead higher-order moments carry the most information. This is denoted as stochastic volatility and in the TBSS literature it is shown that SOBI fails for such time series \cite{Matilainen2017}. Higher-order moments relate often to skewness and kurtosis and, for example in our context, to the covariance of the squared data and are meant to detect more unusual observations (heavy tails). To overcome this issue, a new TBSS method denoted as a variant of SOBI (vSOBI) \cite{Matilainen2017} was introduced. Similar to SOBI, vSOBI finds the latent time series by diagonalizing matrices of lagged fourth moments. Uncovered latent components are uncorrelated marginally and additionally have zero fourth-order dependence.

Generally, time series might carry information both in the autocovariance and in the higher-order time dependence, thus a combination of SOBI and vSOBI might deliver the best results. Indeed, Miettinen et al. \cite{miettinen2020} suggested such a method, referred to as generalized SOBI (gSOBI). It diagonalizes several autocovariance matrices (SOBI part) and several matrices of lagged fourth moments (vSOBI part). This method has three rather involved tuning parameters. The first one $b \in [0,1]$ weighs SOBI versus vSOBI, where SOBI ($b=1$) and vSOBI ($b=0$) are the extreme cases. The second ($\boldsymbol{k}_1$) and third ($\boldsymbol{k}_2$) tuning parameters provide the sets of lags used for the SOBI and vSOBI part, respectively. A lag is a time interval given by a number of time steps, the size of which is determined by the resolution of the underlying time series. For instance, a lag of 6 in an hourly observed thermometer refers to an interval of 6 hours. Common default values for gSOBI are $b=0.9$, $\boldsymbol{k}_1=\{1,\ldots,12\}$ and $\boldsymbol{k}_2=\{1,2,3\}$, but Miettinen et al. \cite{miettinen2020} also show that the selection of lag sets and weight has a huge impact on the performance. Vague guidelines for these tuning parameters exist in the community, such as lag sets should not be too small and not be too large, and the lags should be chosen so that the corresponding (cross-)moment matrices for the latent components have diagonal values far apart. Thus, parameter selection in the context of SOBI is a highly complex problem with no practical solution yet \cite{TANG2005507,TASKINEN201621}.

The R implementation of gSOBI used in the following is available in the package tsBSS \cite{nordhausen2021}. We call one execution of gSOBI a \textit{run}. As outlined before it yields a set of $p$ univariate time series, which we call \textit{components}. The outcomes of multiple runs with varying parameter settings form an ensemble, where each member corresponds to a single run. A member has the used parameters $\boldsymbol{k}_1$, $\boldsymbol{k}_2$ and $b$ associated, as well as the output of gSOBI. The latter is either the component set $\boldsymbol{c}_t$ and the estimated unmixing matrix $\hat{\boldsymbol{W}}$, or nothing, in case the (cross-)moment matrices could not be diagonalized in a predefined number of iterations. We call a run \emph{succeeding} or \emph{failing}, depending on the outcome.

\section{Datasets}
\label{sec:datasets}

In this section we introduce two datasets, one from the medical domain and one from the financial domain, along with reasons why TBSS analysis of them can be desired. Analysis of both datasets shares similar tasks. For instance, analysts are interested in relevant parameter subspaces, common components and alternatives to them, as well as the stability of obtained results. We formalize typical tasks and questions involved in TBSS analysis in \Fref{sec:task-abstraction}.

\begin{figure}[h]
    \centering
     \begin{subfigure}[b]{0.45\textwidth}
         \centering
         \includegraphics[width=\textwidth]{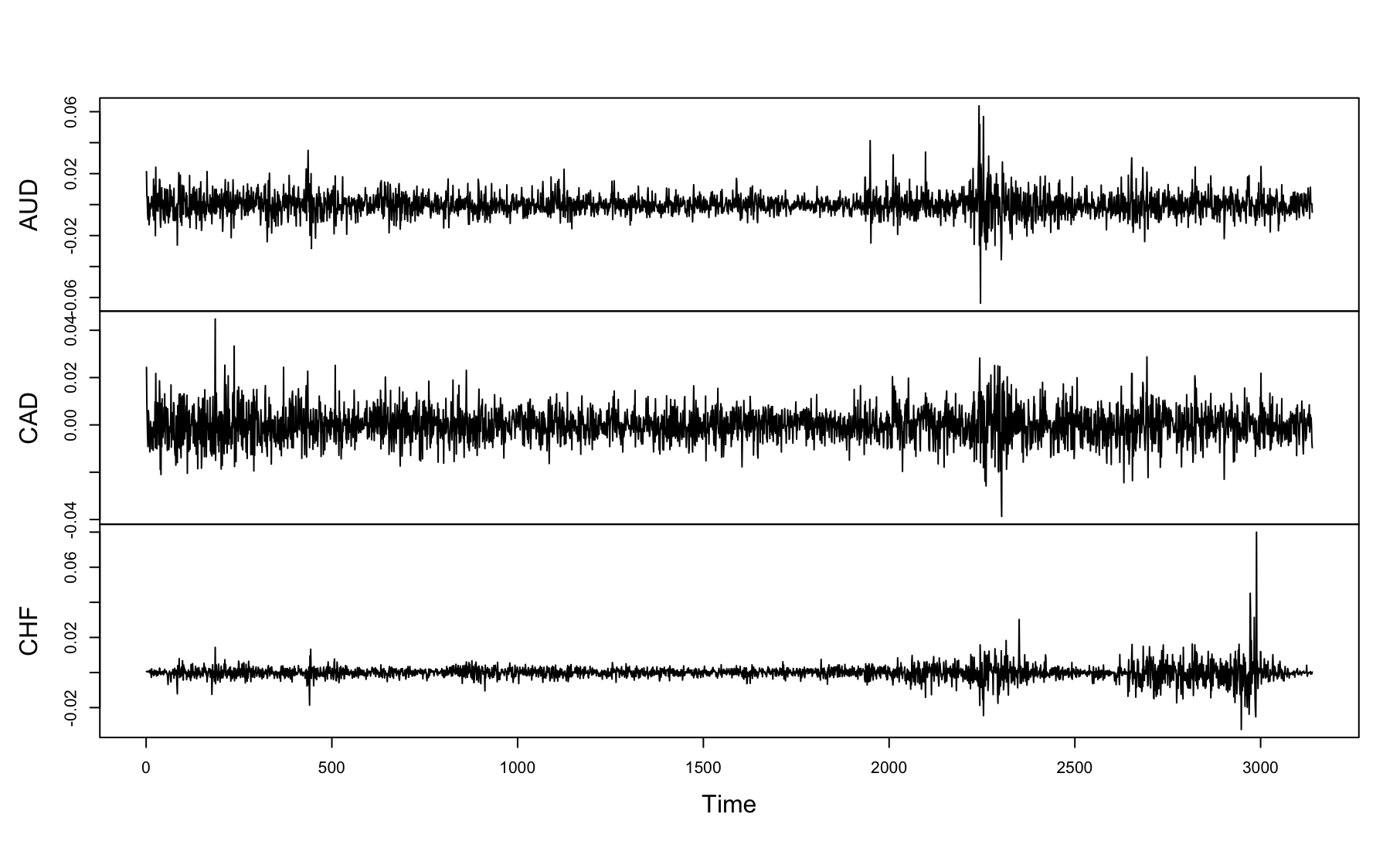}
         \caption{First three currencies of the \emph{exrates} daily currency exchange rate dataset.}
         \label{fig:datasets--exrates}
     \end{subfigure}
     \hfill
     \begin{subfigure}[b]{0.45\textwidth}
         \centering
         \includegraphics[width=\textwidth]{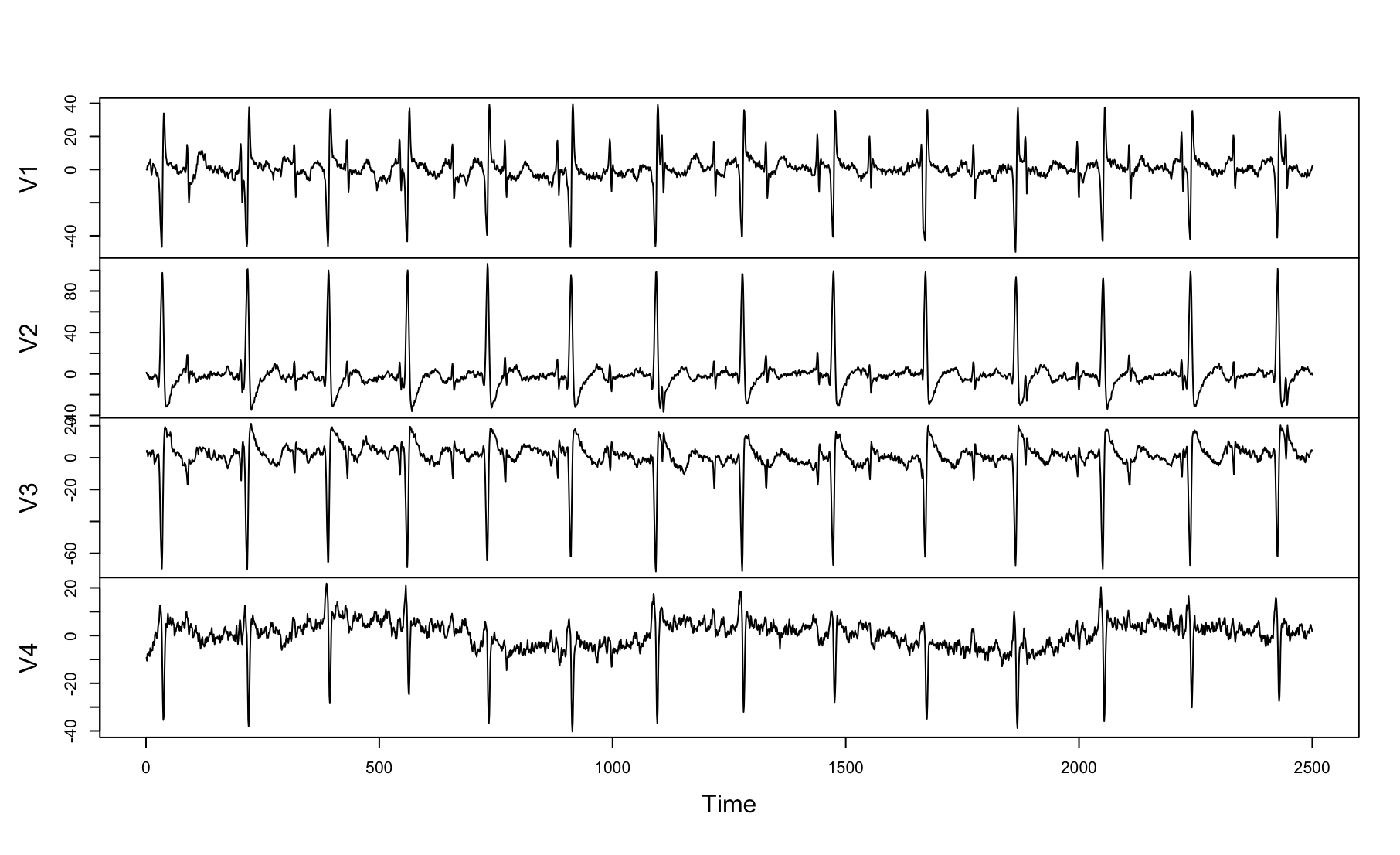}
         \caption{First four variables in the ECG dataset.}
         \label{fig:datasets--ecg}
     \end{subfigure}
     
        \caption{Datasets in this paper.}
        \label{fig:datasets}
\end{figure}

\subsection{Financial data}
\label{sec:datasets--exrates}

Goods, currencies, and company stocks are traded every day at high frequencies. In simple terms, investors make money by buying something at a price X and selling it later at a price Y larger than X. To maximise $Y-X$ in a short time frame the idea here is to find a volatile collection of currencies or stocks (a portfolio), i.e., one that is subject to sudden and extreme changes in value. To do so, we look at the daily exchange rate of 23~currencies to Euro between the years 2000–2012 (23 variables, 3\,139~time steps). We preprocess the data to get logarithmic returns, a common measure in quantitative finance when the temporal behavior of return is of interest. The first three variables are shown in \Fref{fig:datasets--exrates}.

\subsection{Medical data}
\label{sec:datasets--ecg}


An electrocardiogram (ECG) is a recording of the heart's electrical activity. To obtain it, electrodes are placed on the patient's skin. These electrodes detect small electrical changes which occur due to muscle de- and repolarization. ECGs are important for medical analysis as many cardiac abnormalities show deviations to the normal ECG pattern. Analysis of fetal ECGs may detect problems during fetal development, such as fetal distress. While invasive methods exist to measure the fetal ECG directly, a non-invasive method is often preferred as it does not harm neither mother nor fetus. The fetal ECG is visible in the mother's ECG, but it is weak and mixed with, e.g., respiratory noise or frequency interference (compare first three rows in \Fref{fig:datasets--ecg}). Using TBSS on ten seconds of the ECG of a pregnant woman (8~dimensions, 2\,500 time steps), we try to extract the fetal ECG following previous work  \cite{delathauwer2000}.

\section{Task Abstraction}
\label{sec:task-abstraction}

In this section we present a task abstraction for TBSS. We structure it according to the data-users-tasks triangle by Miksch and Aigner \cite{miksch2014} and use the terminology by Brehmer and Munzner \cite{brehmer2013} for tasks. We developed the abstraction together with the visualizations in an iterative design process following Munzner's Nested Model \cite{munzner2009} with three collaborators, who are co-authors of this paper and experts in BSS. In this user-centered design process model, we first conducted unstructured interviews in order to understand their problems and made ourselves familiar with literature they provided. After that, we discussed our assumptions and ideas regularly with them over a course of nine months. We discussed iteratively developed prototypes ranging from hand-drawn sketches, to static digital images, to an interactive application which is described in \Fref{sec:app}. During these sessions, we also questioned our current understanding of their tasks either implicitly through visualization designs or explicitly through discussions. In the end, we interviewed five TBSS experts, who did not collaborate with us on the design, to further validate our abstracted tasks (\Fref{sec:evaluation}). The presented task abstraction is a reflection on this process. 

We touched upon the involved data with TBSS in \Fref{sec:related-work--tbss} already. These are a multivariate time series (input data), one real and two sets of integers (TBSS parameters $b$, $\boldsymbol{k}_1$ and $\boldsymbol{k}_2$) and a set of univariate time series (latent components). The temporal dimension is discrete and linear.

\subsection{Users}
\label{sec:task-abstraction--users}

Our users are data analysts or data scientists with formal education in statistics/math and basic knowledge of BSS. They may also be experts in a specific application domain, like medicine or finance.
They work mostly with R \cite{rcoreteam2020}, a language and environment for statistical computation in which most BSS researchers publish their implementations. The preferred work environment is RStudio, a popular text-based development environment for R. Currently, they use built-in plotting functionality, and sometimes they use, for example, ggplot2 to build customized visualizations. The output of either option is a static visualization, of which RStudio by default displays only one at a time. Because of this, our users are accustomed to well known static statistics visualizations such as histograms, line graphs, box plots, etc.

\subsection{Tasks}
\label{sec:task-abstraction--tasks}

During this user-centered design process we identified the following tasks, which we describe using the abstraction terminology by Brehmer and Munzner \cite{brehmer2013}.

\begin{figure}
  \centering
  \includegraphics[width=\textwidth]{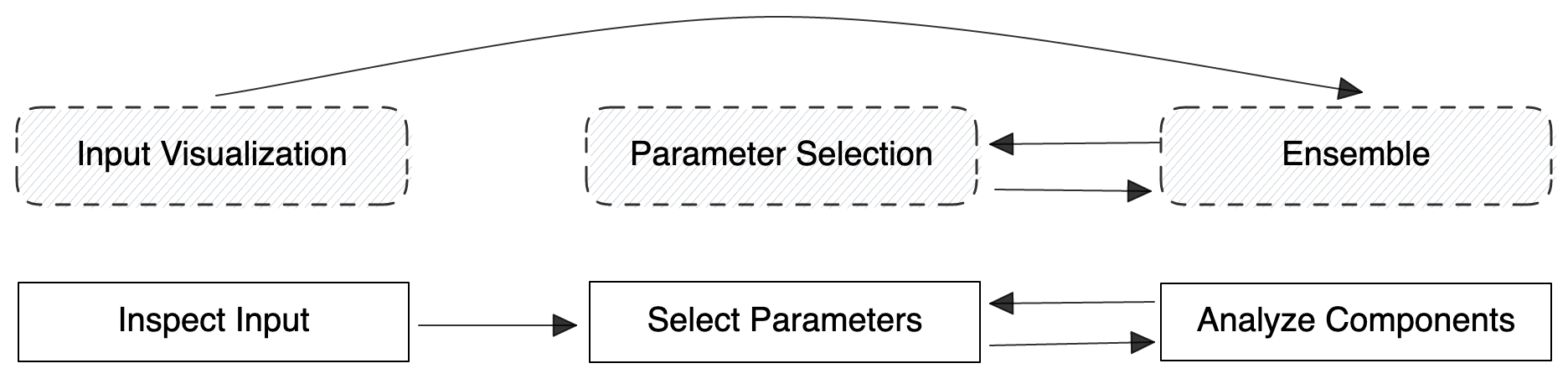}
  \caption{The existing analysis workflow (bottom) and the corresponding screens in \toolname{} (top). The new workflow automatically obtains initial results and analysts can start exploring immediately.}
  \label{fig:app--analysis-phases}
\end{figure}

The high-level workflow can be separated into three phases, which are depicted in \Fref{fig:app--analysis-phases}: Analysts first inspect the raw input data, continue to finding parameter settings, and then analyze obtained components. Given the exploratory nature of their analysis process, analysts switch between the latter two phases until they feel they exhausted the parameter space or obtained a useful result.

Generally analysts want to \textit{discover} observations or \textit{derive} a modified dataset with reduced dimensionality. There are two main \textit{targets} of analysis. One of them are the components, which are mostly analyzed as sets. Still, analysts want to \textit{discover} and \textit{explore} interesting components, whatever interesting means in the data domain. The other analysis targets are the parameters. Analysts look for a \enquote{stable} result, i.e., one that can be obtained with rather diverse parameter settings. The assumption is that its components then more likely represent real processes. To this end, they need to \textit{compare} components and parameters of different runs. As an additional obstacle, when lacking intuition and/or domain knowledge, analysts struggle to select (\textit{lookup}, \textit{locate}) parameters and need guidance support \cite{ceneda2017} so they can \textit{browse} and \textit{explore} parameter settings in an informed way. All these observations lead us to some \textit{low-level query actions}:



\textit{I1: Identify used parameters.} Analysts want to see values of existing parametrizations. In case of lag sets they inspect the distribution of chosen lags and if one lag set contains more lags than the other.

\textit{I2: Identify unmixing matrix.} Analysts turn to the unmixing matrix to interpret components and to understand how they were formed. They look for large absolute values per component.

\textit{I3: Identify cross-moment diagonality.} 
Analysts want to inspect runs on a technical level that is currently inconvenient to obtain and difficult to quantify. If the TBSS model holds, then all cross-moment matrices are exactly diagonalized by the unmixing matrix estimate. For real data, this is, however, rarely the case and thus analysts are interested in the impact of the parameters on the diagonality of the different cross-moment matrices.

\textit{I4: Identify components.} In a single component, analysts look for interesting features like outliers or uncommon changes in shape. They thereby also check the absence of features (noisiness). Analysts are further interested in the stability of a component, i.e., in how many ensemble members the component is present.

\textit{C1: Compare success.} First of all, prior to any comparisons, analysts must know what can be compared. If a run did not succeed, only parameters could be compared as no components were found.

\textit{C2: Compare parameters.} To carry out sensitivity analysis, analysts need to compare parameters between runs. They mainly look at differences in lag distribution and amount.


\textit{C3: Compare unmixing matrices.} Before inspecting factors of individual components, analysts investigate the similarity of unmixing matrixes.

\textit{C4: Compare component sets.} This task mostly relates to membership, which however is difficult to assert with complex objects, such as time series, where one usually speaks of similarity instead of equality. Analysts compare components only between sets and want to know which component exists in both sets, and if so at which ranks, and if not which is the most similar component, plus in which time frames components disagree.

\textit{C5: Compare possible parameters.} When choosing a new parametrization, analysts need guidance through the parameter space and the ability to compare possible parameters in some meaningful way to find a promising setting.

\section{Visualization Design \& Justification}
\label{sec:app}

In this section, we present the visualization design we obtained based on the task abstraction (\Fref{sec:task-abstraction}) and implemented in a web-based prototype for gSOBI. A design goal was to make \toolname{} generic enough to allow its use in many application domains, because, like PCA, TBSS is a domain-independent method. We designed \toolname{} for inputs with the length of up to 5\,000 time steps, and up to 50 dimensions. While these limits do not accommodate extreme cases, like fMRI data (100+ dimensions, 100\,000+ time steps), we expect it is enough for many applications.

While we implemented visualizations for all abstracted tasks, for brevity we will focus on an illustrative subset of those. Specifically, we will discuss visualizations for tasks that pertain to 

\begin{itemize}
    \item identifying and comparing components (or sets thereof),
    \item identifying and comparing used parameters, and
    \item comparing possible parameters.
\end{itemize}

\toolname{} is built of three screens, which are depicted with their connection to analysis phases in \Fref{fig:app--analysis-phases}. The \emph{Input Visualization} screen shows the raw input data, a feature requested by our collaborators. The \emph{Ensemble} screen allows exploration of parameter settings and components. Finally, the \emph{Parameter Selection} screen is used to select new parameter settings. We will focus on the latter two. How presented visualizations work together is illustrated in the usage scenarios (\Fref{sec:usage-scenarios}).

\subsection{Time Series Visualization and Interactions}

\begin{figure}[h]
  \centering
  \includegraphics[width=\columnwidth]{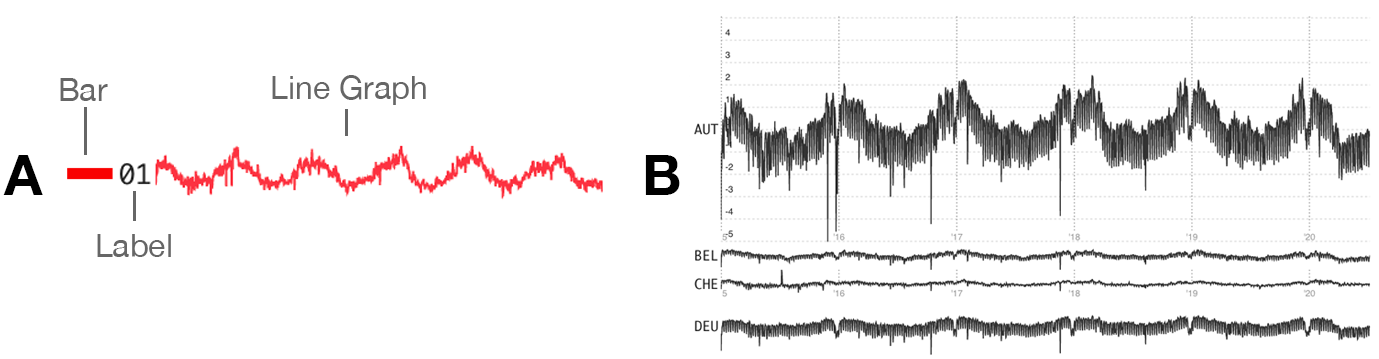}
  \caption{Display of and interaction with time series. (A) A time series is displayed with a line graph, an optional label, and an optional bar to its left. The bar encodes a DOI function of a time series. (B) Typical vertical arrangement of multiple time series in \toolname{}. The user enlarged the first and last time series to different sizes and hovered over the third, thus its X axis labels are shown.}
  \label{fig:time-series-anatomy}
\end{figure}

Time series are plotted vertically aligned to facilitate comparison and ordered by variable name (for input variables) or by an interestingness function (for latent components, see below). The display of and interaction with all time series in \toolname{} is handled by the same logic as shown in \Fref{fig:time-series-anatomy}. Due to the length and amount of time series, we employ semantic zooming and at first save display space by drastically reducing their Y axis and omitting any labels by default. This can be changed with interaction: On hover, we display axis labels for the hovered time series. The Y axis can be increased individually by another interaction. If an analyst is interested in a contiguous subset of the time series, it is possible to zoom in with brushing, which will affect all time series in the application. Both the semantic and temporal zoom can be reset with interactions recommended by Schwab et al. \cite{schwab2019}.

As described in \Fref{sec:related-work--tbss}, the order of components is not defined. In practice, this means that analysts use measures which are sign-independent to compare components, such as absolute Pearson correlation, and impose an order by sorting components according to a function. We will call this a \textit{degree-of-interestingness function} (DOI), and require it to be any function $f: \mathbb{R}^n \rightarrow \mathbb{R}$ that maps a time series of length $n$ to a single number. Because TBSS is a domain-independent method, many DOI functions could be useful \cite{fu2011} depending on what the domain's interesting features are. E.g., for detailed cardiac analysis, different widths and types of ECG wave patterns could be mined. Based on discussions with our collaborators we use the absolute third (skewness) and fourth (kurtosis) moment in \toolname{}. These are useful to find the most skewed components and those with the most outlying values, as the first two moments (mean and variation) of all components are identical. We added a measure for periodicity \cite{vlachos2005} after our user studies. The DOI function can be changed in the toolbar, and all views that show component-related data will update as the set members are sorted in descending order based on the new DOI function values.

\subsection{Color}

According to Mackinlay \cite{mackinlay1986}, color is the most effective visual variable for nominal data after position, and, therefore, often used to encode different data classes. In multiple views, the same classes should be encoded with the same palette \cite{qu2018}. Because humans can only reasonably distinguish a few different colors, we cannot statically assign colors to all ensemble members. We, therefore, use a user-controlled dynamic assignment of colors of a qualitative palette to encode data related to user-selected members. The available colors are displayed in the toolbar and can be reordered with drag \& drop. When hovering over an unselected member, the next free color (left to right) is a) used to highlight its related data in all views and b) associated to the member when selected. As one color is always needed for highlighting, the last free color cannot be used for selection. The color order determines the plotting order in all comparison views.

\subsection{Tasks I4/C4: Identify/Compare Components (Sets)}
\label{sec:app--compare-comonents}

We precompute initial parameter settings automatically, to allow immediate exploration of the result space: Variations of gSOBI's R package's defaults (3 settings), a recommendation from the literature \cite{tang2005a} (1 setting), and an additional user-defined number of random settings. This overcomes initial hesitation towards parameter setting choice and may give an estimate of what the relevant parameter subspace is.

Each successful run (\Fref{sec:related-work--tbss}) produces a set of time series. Already at the start of the analysis after precomputation, the amount of components to consider might be in the hundreds. Clustering is an established approach to counteract this, where data cases are grouped by similarity. This allows an analyst to focus on representative elements of the clusters. Many clustering techniques exist \cite{xu2005a, xu2015}, but using them with all components as they are has a major drawback: The clustering scheme will group components from the same set, which our collaborators found undesirable. The grouping should respect the set structure in the data and group components only between sets, not within them. Additional requirements we gathered for the clustering scheme are that it should not depend on a distance metric (unlike, e.g., k-means) and produce an existing data case as cluster representative (again unlike, e.g., k-means). The former is related to the similarity measure for components suggested by our collaborators, the difference in absolute Pearson correlation $dist_{cor} = 1 - | cor(c_i, c_j) |$. Since we do not know if it supports the triangle inequality, we should not rely on it. The latter requirement stems from the design principle to show actual data over visual abstractions.

We developed a custom clustering scheme to achieve these requirements. Starting from the realization that we basically want k-medoids as it does not need a distance and produces existing representatives (medoids), we looked for a way to constrain the clustering process to obey the set structure. Constrained versions exist for k-means \cite{wagstaff2001}, but we did not find one for k-medoids. However, it was possible to adapt it using a k-means-like formulation of k-medoids \cite{park2009}. Constraints in our case are of the type \emph{cannot-link}, i.e., they express which data cases must not be grouped into the same cluster. We add one cannot-link constraint per pair of elements that belong to the same set. For $m$ sets containing $p$ data cases each this amounts to $mp(p-1)/2$ constraints in total. Algorithm~\ref{alg:cop-pam} shows pseudocode of our custom clustering scheme.

\begin{algorithm}
\DontPrintSemicolon 
\caption{Pseudocode of constrained k-medoids, with which we obtain a clustering on sets of time series. The function FastPAM refers to a recent k-medoids algorithm without constraints \cite{schubert2019}.}
\label{alg:cop-pam}
\KwData{Dissimilarity Matrix $D$, constraints $C$, medoids $M$, assignments of data cases to medoids $A$, number of partitions $k$}
$M, A \gets FastPAM(D, k)$\;
$cost \gets getCost(A, D)$\;
$cost' \gets cost$\;
\While{$violatesConstraint(A, C)$ or $cost - cost' > \epsilon$}{
 $M' \gets findNewMedoids(A, D, k)$\;
 $A' \gets assignToNearestPossibleMedoid(M', C, D)$\;
 \;
 $cost \gets cost'$\;
 $cost' \gets getCost(A', D)$\;
 $A \gets A'$\;
 $M \gets M'$\;
}
\KwResult{$(M, A)$}
\;
\end{algorithm}

We show the clustering result to the analyst with the following visualizations.

\subsubsection{Clustering Quality and Number of Partitions}
\label{sec:app--compare-components--clustering-quality}

The constrained k-medoids clustering takes one user-provided parameter, which is the desired number of partitions. We use a scented widget \cite{willett2007} to allow setting this parameter in an informed way (\Fref{fig:app--focus--components}, A). The bar chart in the widget shows the average cluster separation as a clustering quality measure for a given number of clusters. Therefore, values with high bars suggest the number of meaningfully different components in all currently available sets.

\subsubsection{Component Overview}
\label{sec:app--compare-components--component-overview}

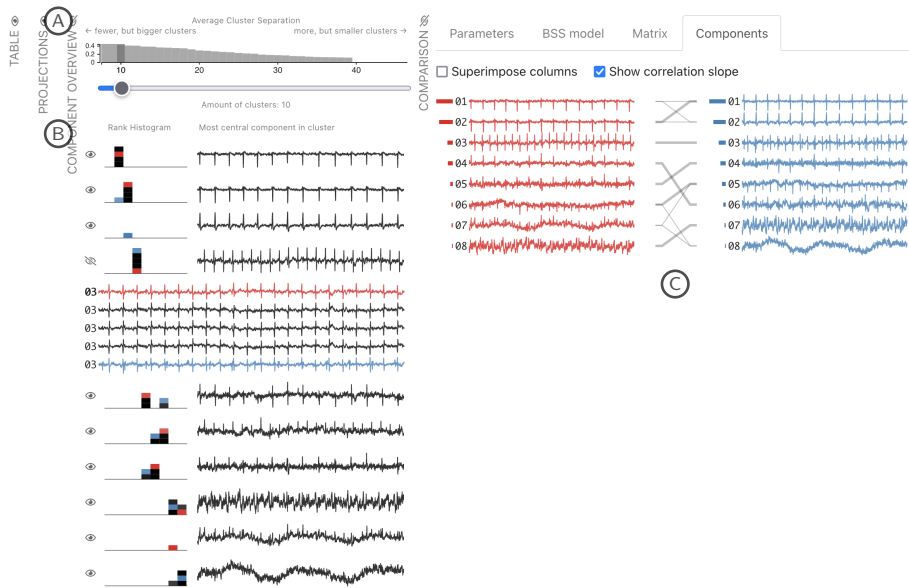
\begin{figure}
    \centering
    \begin{tikzpicture}
	\node[anchor=south west,inner sep=0] (image) at (0,0)  {\includegraphics[width=\textwidth]{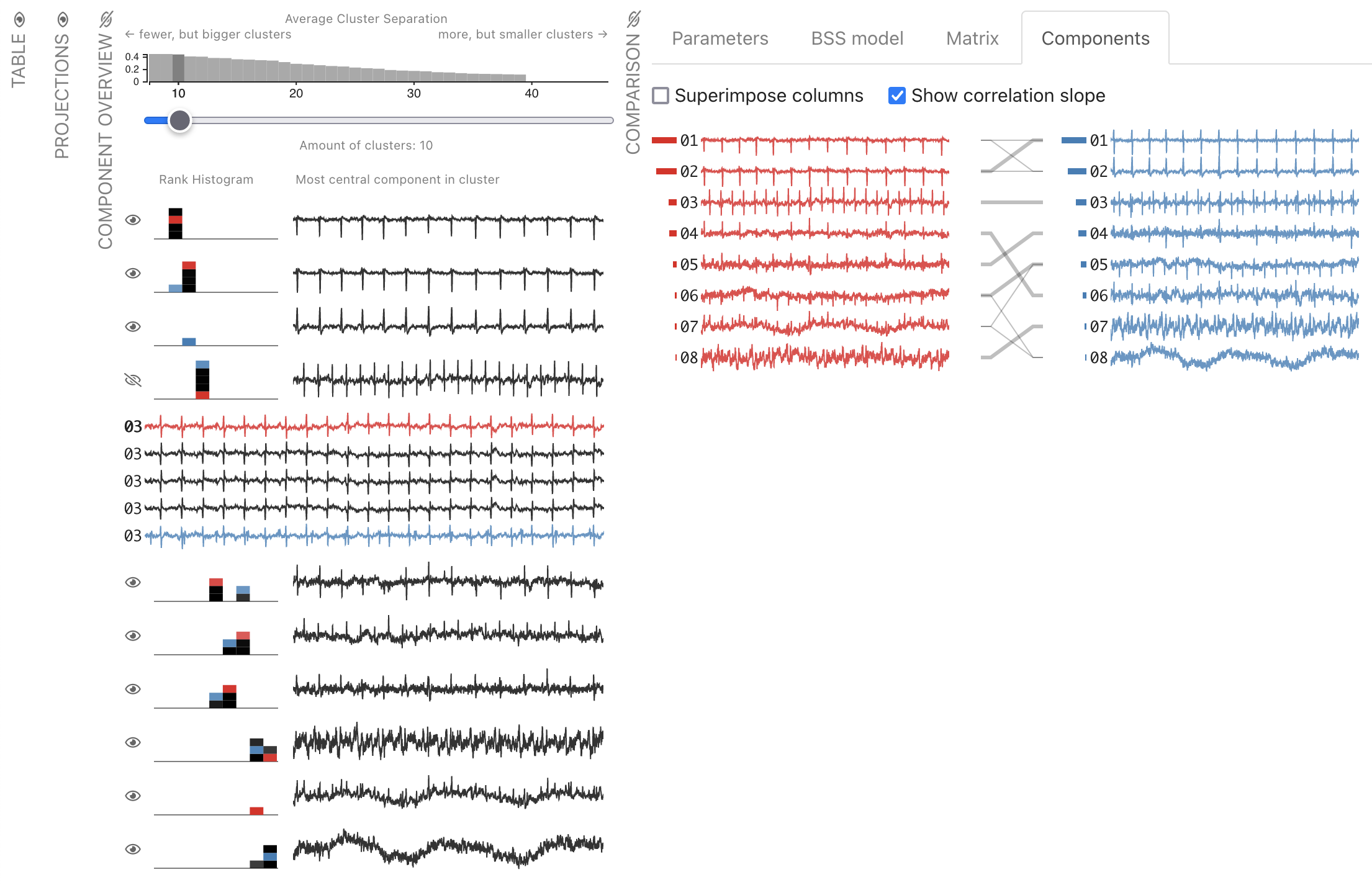}};
    \begin{scope}[x={(image.north west)},y={(image.south east)}]
		\node [textWcolor, anchor=north west] (a) at (0.995, 0.05) {A} ;
		\node [textWcolor] (b) at ($(a)-(0mm, 15mm)$) {B} ;
		\node [textWcolor] (c) at ($(b)+(82mm, -20mm)$) {C} ;
    \end{scope}
    \end{tikzpicture}
    \caption{\emph{Ensemble} screen of \toolname{} (ECG data) configured to facilitate comparison and inspection of components and sets thereof. Left part shows the Clustering Quality (A), which suggests an optimal clustering with 8--10 partitions. Medoids of the 10 clusters are listed underneath (B). The fourth list item readily shows the fetal heart signal and was expanded to show cluster members. Two component sets (red and blue) were selected for detailed comparison (right half). The Slope Graph (C) highlights similar components, their rank changes and set similarity overall.}
    \label{fig:app--focus--components}
\end{figure}

The cluster medoids are shown underneath vertically aligned in a list, sorted by the DOI rank of the medoid (\Fref{fig:app--focus--components}, B). To further support Task C4, we show a histogram to the left of the medoid. The histogram shows the rank distribution of the contained components in their respective sets. Additionally, we encode $dist_{cor}$ to the cluster medoid with opacity. This way, stable (stacked bars with high opacity) and unstable (scattered bars with low opacity) components have distinct histogram shapes. 

Analysts can inspect components in a cluster by clicking the ``eye'' icon, after which the list item expands and lists contained components in the same fashion as cluster medoids. Clicking a bar in the histogram or a time series label selects the associated ensemble member.

\subsubsection{Slope Graph}
\label{sec:app--compare-components--slope-gaph}

Components of selected sets are visible in a separate view, again vertically aligned and sorted by DOI (\Fref{fig:app--focus--components}, C). Each selected set has a unique assigned color and all associated data is shown in this color. Multiple selections are juxtaposed horizontally in columns, which can be rearranged by the analyst. Analysts can inspect components visually as they are, or they can also display a slope graph between columns. Lines of the slope graph connect similar components, and thickness encodes similarity from high correlation (thick) to low. This way, it is easy to see stable (thick, single, mostly straight lines) and unstable components (no or thin, multiple, tilted lines), their rank changes and set similarity at a glance. 

\subsection{Tasks I1/C2: Identify/Compare Used Parameters}
\label{sec:app--compare-parameters}

\begin{figure}
    \centering
    \begin{tikzpicture}
	\node[anchor=south west,inner sep=0] (image) at (0,0)  {\includegraphics[width=\textwidth]{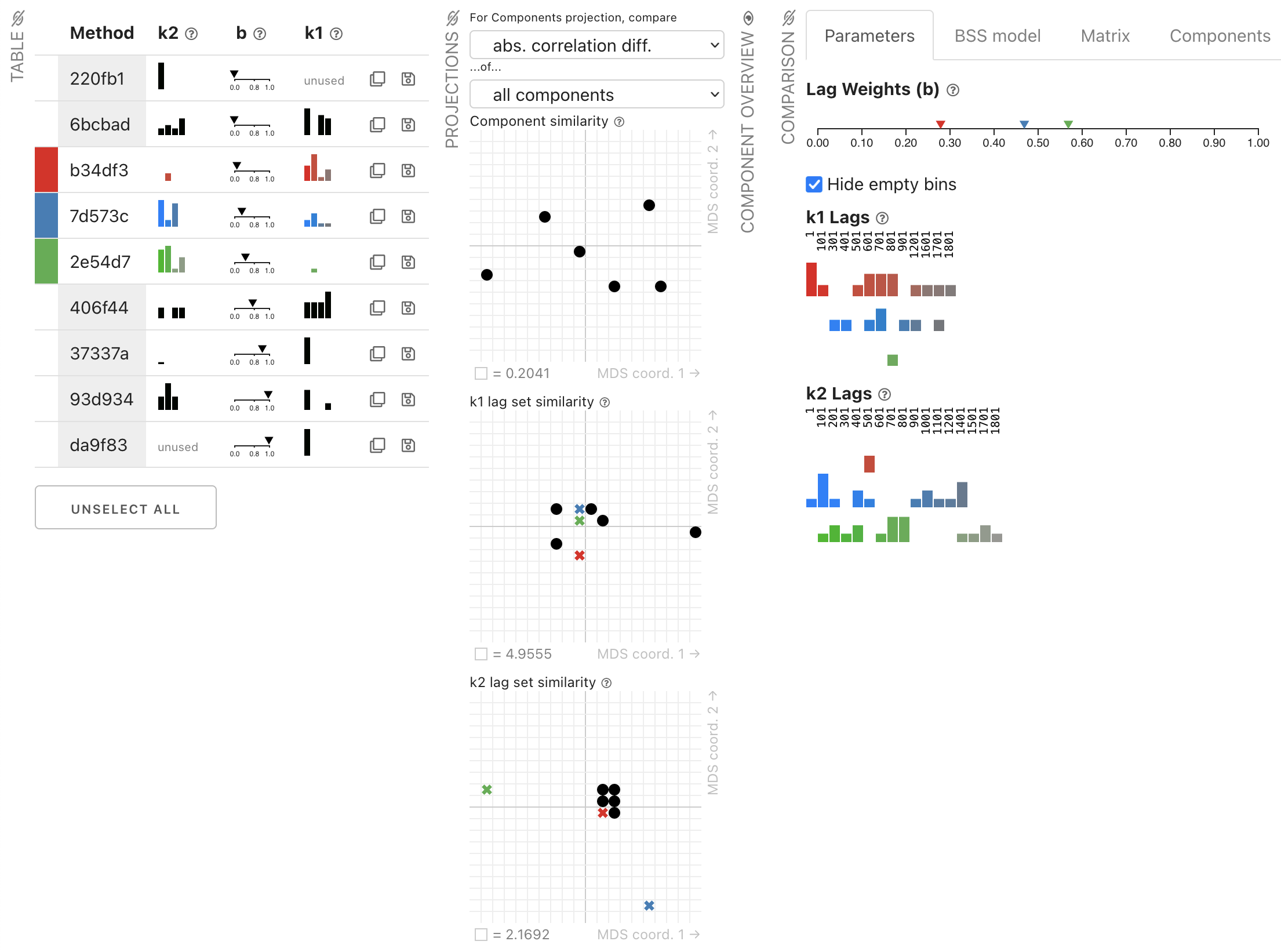}};
    \begin{scope}[x={(image.north west)},y={(image.south east)}]
		\node [textWcolor, anchor=north west] (a) at (0.9, 0) {A} ;
		\node [textWcolor] (b) at ($(a)+(40mm, 0mm)$) {B} ;
		\node [textWcolor] (c) at ($(b)+(32mm, 0mm)$) {C} ;
    \end{scope}
    \end{tikzpicture}
    \caption{\emph{Ensemble} screen of \toolname{} (ECG data) configured to facilitate comparison and inspection of parameters. Three failing runs were selected. Left column shows a tabular overview (A). Middle column shows DR projections of component and parameter similarities (B). Right column shows detailed comparison views to facilitate parameter comparison (C). It is apparent that failing runs had a weight parameter $b$ of 0.25--0.6 and $\boldsymbol{k}_1$/$\boldsymbol{k}_2$ lag sets that span the whole range, which suggests that this parameter subspace should be avoided.}
    \label{fig:app--focus--parameters}
\end{figure}
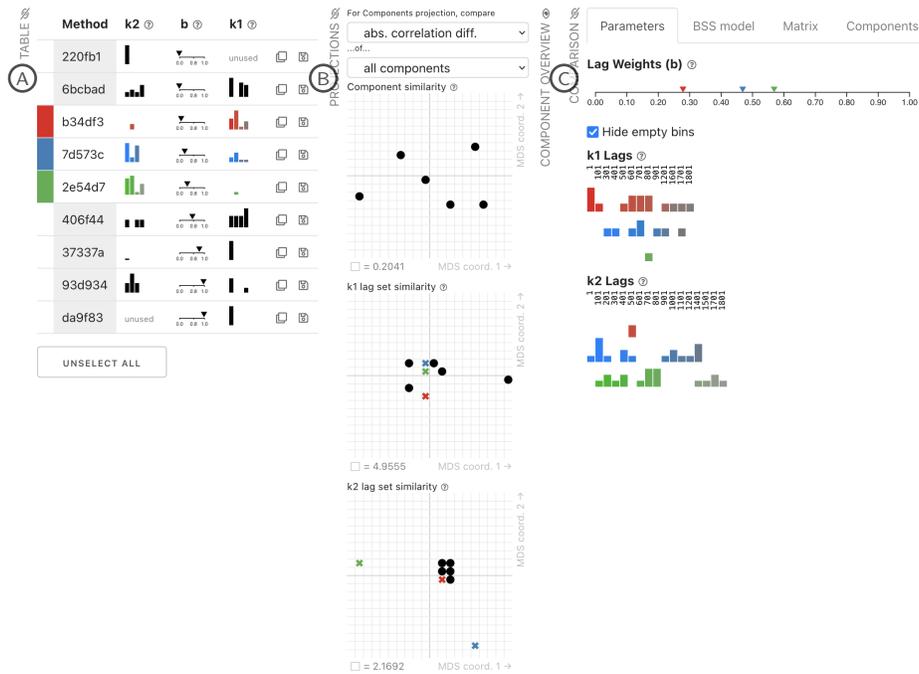

Parameter space analysis \cite{sedlmair2014} is another important task for BSS experts, where they are mainly interested in \emph{sensitivity analysis} and \emph{partitioning}. We facilitate these tasks with tailored visualizations (\Fref{fig:app--focus--parameters}).

\subsubsection{Similarity Views}
\label{sec:app--compare-parameters--similarity-views}

Similarity of so far obtained component sets, as well as selected parameters, are shown in three separate dimensionally-reduced views. Marks that are close to each other suggest similar components and  $\boldsymbol{k}_1$/$\boldsymbol{k}_2$ parameters. Multidimensional Scaling (MDS) is an appropriate dimension reduction technique for global cluster analysis according to recent publications \cite{nonato2019, xia2022}. We use non-metric MDS \cite{venables2010} as we do not always have a distance metric. As MDS will project elements with same values in high-dimensional space to the same low-dimensional points, we would soon run into an occlusion problem—consider an analyst who keeps lag sets the same, but varies only the weight. There are a couple of ways to deal with occlusion, most notably lenses \cite{tominski2017}. However, our users are not used to complex interactions, so we changed the tradeoff between position accuracy and occlusion. As an implementation of CorrelatedMultiples \cite{liu2018a} was not available, we only rasterize the MDS plot and move overlapping points to the next free cell. When hovering over a point, the other points will change their size proportionally to the original dissimilarity, thereby allowing analysts to investigate projection errors.

\subsubsection{Parameter Comparison} 
\label{sec:app--compare-parameters--parameter-comparison}

\begin{figure}
    \centering
    \includegraphics[width=0.75\linewidth]{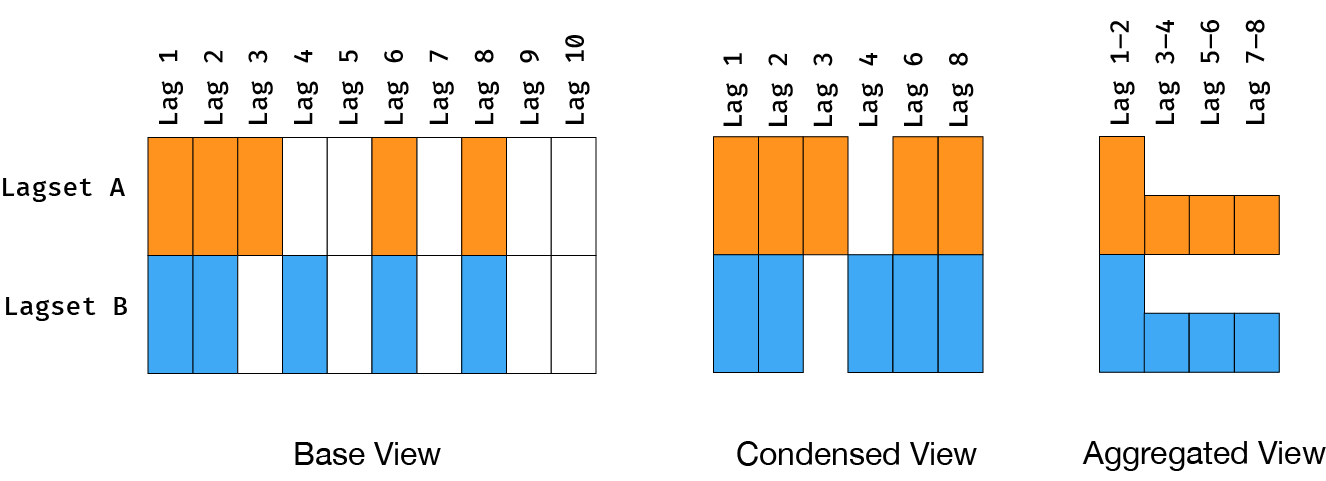}
    \caption{Interweaved histograms encoding two lag sets A and B facilitate comparison. The base version encodes the presence of a lag by filling the corresponding rectangle with color hue. The condensed version, which is the default, omits lags that are in no lag set. Finally, when increasing the bin size (pictured: to 2), the analyst sees the aggregated view, where rectangles are transformed in small bar charts.}
    \label{fig:app--interweaved-lags}
\end{figure}

To compare weights of different parametrizations, we encode triangle marks on a shared axis. Triangles are stacked in case they otherwise completely occlude each other. To compare lag sets, we use interweaved histograms where the color saturation of a bar encodes the lag size to give an additional visual hint of the lag distribution, and to be consistent with the encoding in the lag selection (\Fref{sec:app--parameter-selection--lag-selection}). \Fref{fig:app--interweaved-lags} shows how they are generated. First, individual bars are positioned in a grid such that bars of the same lag set are in the same row, and bars representing the same bin are in the same column (base view). To save display space, empty columns are hidden by default (condensed view), but can be shown after user interaction. Increasing the bin size leads to familiar histogram shapes (aggregated view). Interweaved histograms show distinct images for same (bars align vertically and have similar height) and different lag sets (bars appear interweaved).

\subsection{Task C5: Compare Possible Parameters}

To obtain a new result, analysts need to select parameters. They consist in the case of gSOBI of two lag sets and one weight (\Fref{sec:related-work--tbss}).

To facilitate this selection process, we used the guidance design framework \cite{ceneda2020} to design appropriate guidance \cite{ceneda2017}. Analysts do not know which lags to select and are generally aware of this \emph{knowledge gap}. As discussed in \Fref{sec:task-abstraction--tasks}, the \emph{analysis goal} is to obtain a new/interesting result. Issues occur in the phase of lag selection, because the space of possible lag sets is huge. Analysts currently do not use additional information about lags, mostly due to time constraints. The \emph{knowledge gap} lies in the execution and relates to the input data. We opt for \emph{orienting} guidance, because analysts select lags also based on past experience and domain knowledge, so stronger guidance could be detrimental, and because our guidance \emph{input} is not (cannot be) the ``true'' data: We compute it from the input data, which are per BSS model a linear combination of the components we are interested in. Based on the input data, we calculate guidance \emph{output} per lag that help relate them to each other:

\textit{Guidance Output (GO) 1: Calendar relation.} We compute which lag fits best to intervals in bigger calendar granules. The benefit of this is two-fold. First, lags are abstract and do not consider the calendar used in the data, so thinking in terms of days, weeks, etc., is a more intuitive alternative for someone familiar with the data. Second, it allows us to organize lags by filtering to those which correspond to a difference in a given calendar granule, thereby reducing the amount of lags to reason about. 

\textit{GO2: Largest autocorrelation in input time series.} White noise is a serially uncorrelated process, i.e., does not exhibit autocorrelation, so this measure indicates a latent component might not be white noise.

\textit{GO3: Eigenvalue difference in autocovariance matrices.} The analysts use it to learn more about the input data and it can inform the parameter selection as lags should be chosen such that this eigenvalue difference is big (see \Fref{sec:related-work--tbss}).

\textit{GO4: Cross-moment matrix diagonality.} This can only be computed when a parametrization of a successful run is refined, i.e., an unmixing matrix estimate exists. It shows the analyst which selected lags had an impact on the diagonality of autocovariance and fourth cross-cumulant matrices. It can be understood as feedback into the guidance system. 

\subsubsection{Lag Selection}
\label{sec:app--parameter-selection--lag-selection}

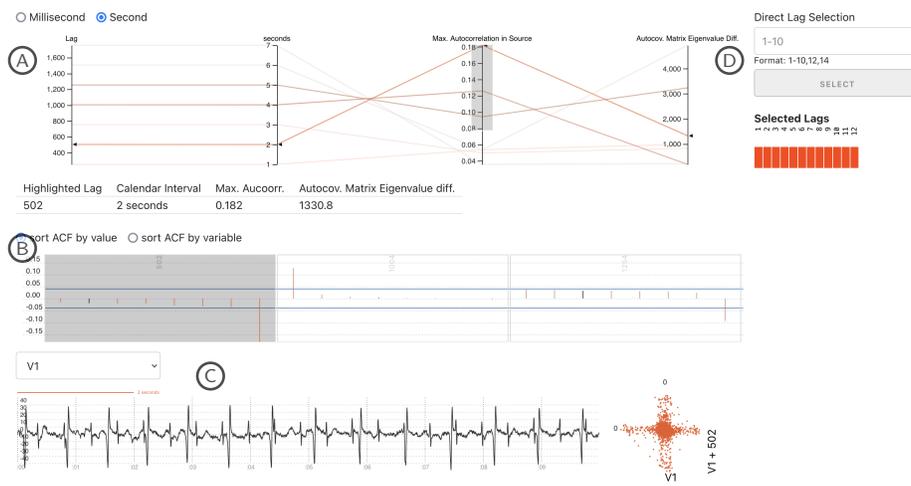
\begin{figure}
  \centering
  \begin{tikzpicture}
	\node[anchor=south west,inner sep=0] (image) at (0,0)  {\includegraphics[width=\textwidth]{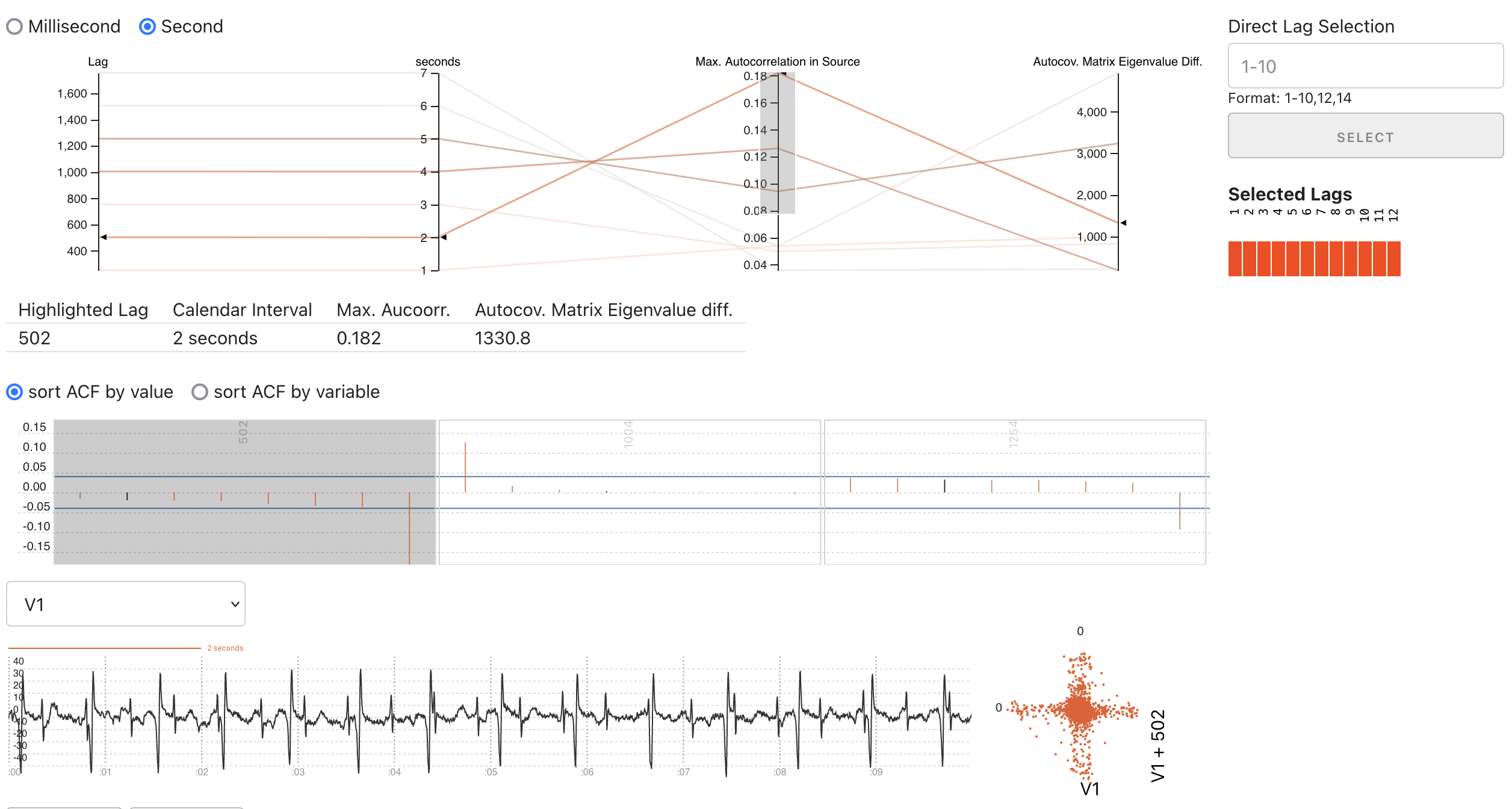}};
    \begin{scope}[x={(image.north west)},y={(image.south east)}]
		\node [textWcolor, anchor=north west] (a) at (0.9, 0.0) {A} ;
		\node [textWcolor] (b) at ($(a)+(0mm, -25mm)$) {B} ;
		\node [textWcolor] (c) at ($(a)+(25mm, -42mm)$) {C} ;
		\node [textWcolor] (d) at ($(a)+(94mm, 0mm)$) {D} ;
    \end{scope}
    \end{tikzpicture}
  \caption{The Lag Selection view (ECG data): Lag size is encoded with color saturation. Lags are filtered to those corresponding to a temporal difference of multiple seconds in the underlying calendar. The PCP (A) further narrows them down to those with high autocorrelation. A multivariate ACF plot (B) shows the autocorrelation of input data at brushed lags. Lags can be selected by clicking and highlighted by hovering in the ACF plot. A user-selected input time series is shown underneath (C) next to a scatterplot of the datapoints of the series at the currently highlighted lag. The right-most column (D) allows analysts to skip the interaction, and shows the current selection.}
  \label{fig:app--lag-selection}
\end{figure}

We support selection of a single lag set with multiple coordinated views (see \Fref{fig:app--lag-selection}). The lag size is encoded with color saturation, to make long, medium, and short lags distinguishable in all views, which is roughly how analysts reason about lag sets.

A parallel coordinates plot (PCP) displays all lags corresponding to a selected calendar granule (\Fref{fig:app--lag-selection}, A), which can be configured by the user. Its dimensions are GO1–4, and values of selected lags are displayed as triangle marks next to the axes. The PCP supports common interactions such as inverting dimensions, reordering dimensions, and brushing.
It is used to reduce the parameter space to a manageable subset. This subset is then visualized in a multivariate autocorrelation function plot (MACF), so analysts can view the temporal structure of all variables (\Fref{fig:app--lag-selection}, B). The MACF shows all univariate autocorrelation function plots, composited through nesting: One box contains the autocorrelations of all variables at a given lag. Autocorrelations are encoded as bars, as in the univariate version, and can be sorted by variable name or by value. The latter is the default because it shows the distribution of autocorrelation values. Hovering over a box highlights the lag, which affects the next view below it. Clicking a box adds or removes the lag from/to the selection, which is shown in the right column in the same fashion as interweaved histograms (\Fref{fig:app--interweaved-lags}).

Underneath the MACF we display a user-selected input time series as line graph, and a scatterplot of the time series' values vs. the values lagged by the currently highlighted lag (\Fref{fig:app--lag-selection}, C). This allows the analyst to find correlation patterns which are not surfaced by the MACF or the time series itself. A line on top of the time series shows the extent of the currently highlighted lag in context. 

These views allow the analyst to interactively explore possible lags. Should they exactly know what they want to select, or rather not use an interactive system because they are used to static tools, they can enter the desired lags in the input box in the right column (\Fref{fig:app--lag-selection}, D) in a format similar to R's \texttt{seq} shorthand syntax and proceed.

\subsection{Task C3/I2: Compare Unmixing Matrices}
\label{sec:app--compare-unmixing-matrix}


We support this task (I2) by showing the factors as a heatmap where a univariate color scale encodes the absolute value in a row with white (low value) to black. When analysts see interesting patterns, they can select cells, and the respective input data and components will be shown underneath the matrices (\Fref{fig:app--matrix--exrates}). This allows to investigate the relationship between inputs and components. Task C3 is also supported, for which we encode a BSS-specific similarity measure \cite{ilmonen2010} in a heatmap with a univariate color scale.

\section{Usage Scenarios}
\label{sec:usage-scenarios}

In this section we describe how the designed visualizations (\Fref{sec:app}) allow insights into the presented datasets (\Fref{sec:datasets}). The financial dataset was used in our user studies (\Fref{sec:evaluation}), while we added the medical dataset ourselves to provide broader context to the reader. The usage scenarios we describe are based on what we learned during aforementioned user studies and also during discussions with our collaborators.

\subsection{Financial data}

\begin{figure}
    \centering
    \includegraphics[width=0.5\linewidth]{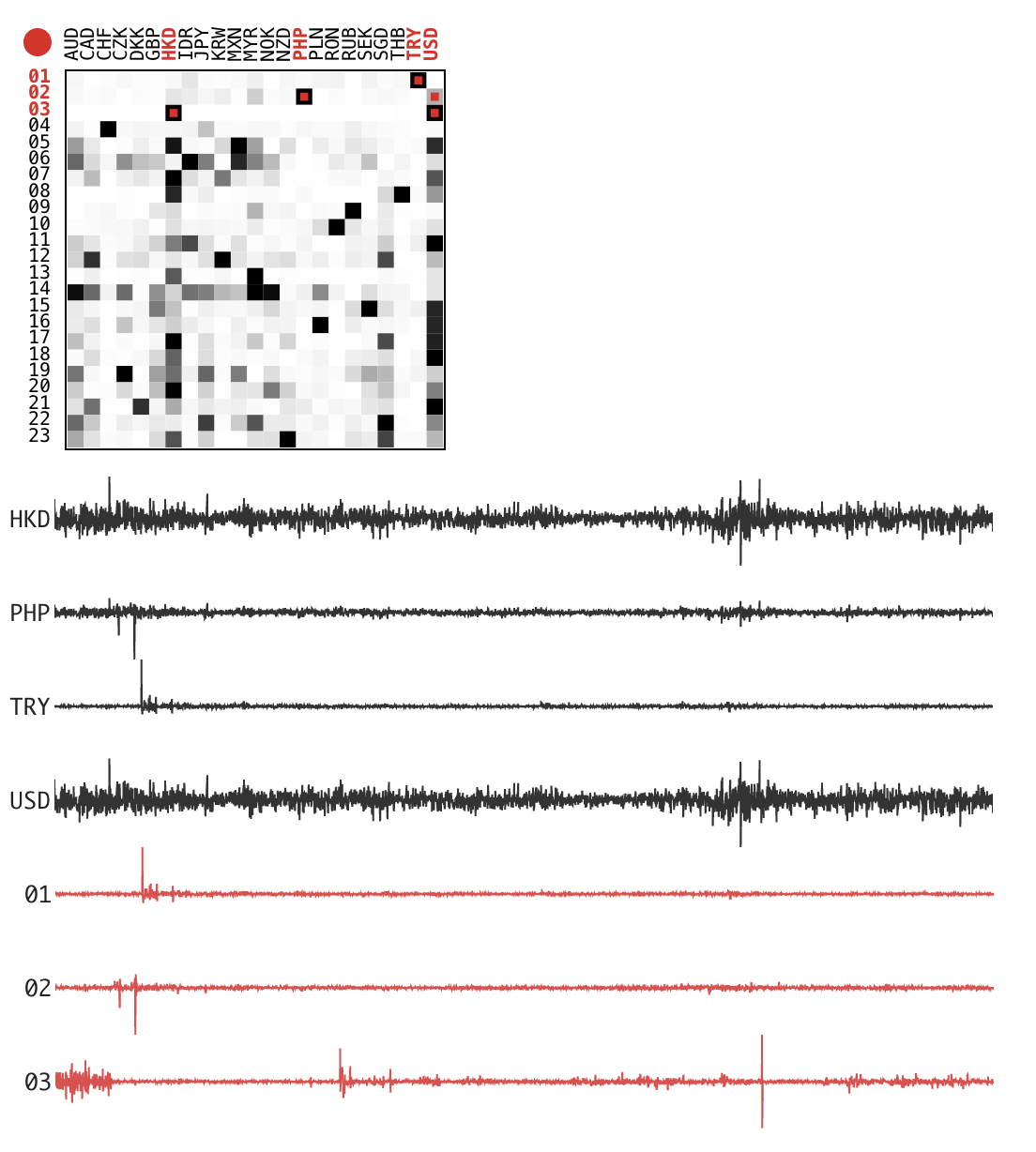}
    \caption{Unmixing matrix visualization from the \emph{exrates} dataset. The most influencing variables (black) for the first three components (red) were selected.}
    \label{fig:app--matrix--exrates}
\end{figure}

We load the financial dataset (\Fref{sec:datasets--exrates}) of 23 currency exchange rates to Euro into \toolname{} and start with 9 parameter settings. From the \emph{Component Similarity View} (\Fref{sec:app--compare-parameters--similarity-views}) we can immediately see that two component sets are very similar. Selecting them reveals that one of them did not use the $\boldsymbol{k}_1$ lag set at all, suggesting that this parameter's influence is small and we should focus on $\boldsymbol{k}_2$ when selecting parameters. This is in line with our expectations of financial data (\Fref{sec:related-work--tbss}). Looking at the clustering visualizations, no clear picture emerges. The \emph{Clustering Quality} (\Fref{sec:app--compare-components--clustering-quality}) increases slowly with the number of clusters, but there is no distinctive peak. Inspection of the \emph{Component Overview} (\Fref{sec:app--compare-components--component-overview}) confirms that many components share a similar pattern, they are very noisy with more extreme values during the years 2008/2009. This was the time of the global financial crisis. Given that nothing really stands out here, we try to obtain an alternative result and go to the \emph{Parameter Selection}. We set the weight $b$ to zero and do not use SOBI part ($\boldsymbol{k}_1$ parameter) at all, following our initial hypothesis. In the \emph{Lag Selection} (\Fref{sec:app--parameter-selection--lag-selection}) for $\boldsymbol{k}_2$ we quickly select lags that correspond to 1–3 days, 1–4 weeks, 1–3 months and 1 year intervals in the underlying calendar. We do it this way because the other guidance outputs do not seem informative due to the amount of noise in the dataset. The newly computed result is colored green in \toolname{} and automatically selected. We look at its components and compare it to the two identical results. The \emph{Slope Graph} (\Fref{sec:app--compare-components--slope-gaph}) shows many thick lines that connect identical components. As we want to find currencies to invest in, we turn to the \emph{Component Overview} again. The histograms suggest that the first couple of components are common in all results, i.e., are stable. We therefore pick three that have volatile segments outside of 2008/2009 to rule out a global financial crisis as the cause for volatility. The \emph{Unmixing Matrix} visualization (\Fref{sec:app--compare-unmixing-matrix}) shows which currencies are associated with these components (\Fref{fig:app--matrix--exrates}). We will ask our financial advisor about investing in Thai bhat, US dollars, Turkish lira, or Philippine pesos. 

\subsection{Medical data}

We load the ECG dataset (\Fref{sec:datasets--ecg}) from a pregnant woman into \toolname{}. Looking at the raw inputs in the \emph{Input Visualization} we can confirm that the fetal heart signal is visible in the mother's ECG. We start with 9 precomputed parameter settings, 6 of which succeed. The \emph{Clustering Quality} (\Fref{sec:app--compare-components--clustering-quality}) suggests that 8--10 meaningfully different components were obtained (\Fref{fig:app--focus--components}, A). We set the clustering to 10 partitions. A healthy fetus has a heart rate of 110--160 beats/minute on average, which is higher than that of an adult (60--100). A candidate component for the fetal heart signal, which shows peaks of increased frequency, is readily visible as 4th in the \emph{Component Overview} (\Fref{sec:app--compare-components--component-overview}). The rank histogram next to the cluster medoid shows that components in the cluster are very similar, which is confirmed by looking at them directly (\Fref{fig:app--focus--components}, B). We select a couple of results containing this component to compare their parameters. In  \Fref{sec:app--compare-parameters--parameter-comparison} we see that the parameters vary wildly, and the fetal heart signal was found using long and short lags for either lag set with different weights. This, along with the absence of other candidate components, suggests that we found the correct signal. A medical doctor would be able to inspect the obtained fetal ECG wave patterns in detail and determine whether or not it is healthy.

Looking at the values of the three parameter settings that did not produce results, we can also form an initial hypothesis about the useful parameter subspace (\Fref{fig:app--focus--parameters}). What they had in common was i) a weight $b$ between 0.25 and 0.6 and ii) lags that were distributed over the whole range instead of sticking to either the short or long end. Thus, when trying to find new parameters for this dataset, we would steer clear of those properties.


\section{Evaluation}
\label{sec:evaluation}

To assess the usefulness of our visualization design, we conducted two interviews with five TBSS experts external to the project. Our research questions were:

\begin{itemize}
    \item[RQ1] What are advantages and disadvantages of \toolname{} in comparison to their current tools?
    \item[RQ2] Does \toolname{} in fact support the analysis tasks?
    \item[RQ3] What are possible improvements to \toolname{}?
\end{itemize}

We decided for an \emph{Expert Review} \cite{elmqvist2015} using interviews, as no comparable tool for a quantitative evaluation exists and qualitative data allows much deeper insights. Two interview cycles were conducted: The first to gather initial external feedback and supporting evidence for our task abstraction, and the second to verify that this feedback was integrated accordingly. They lasted 2.5 hours and 1 hour, respectively.

\subsection{Participants}
\label{sec:evaluation--participants}

Participants were the same for both interviews and previous collaborators of our co-authors. They participated voluntarily without promised benefits, financial or other. All are adults and not dependent on any author, be it financially, professionally or personally.

Our five experts (E1–E5) all hold a Ph.D. degree in mathematics or statistics. Four obtained their Ph.D. with research in BSS somewhat recently, while the other researches BSS already for 10~years. Therefore, they more than fit the \emph{basic knowledge of BSS} and \emph{formal education in math/statistics} requirements from our users characterization (\Fref{sec:task-abstraction--users}). Since participants applied (T)BSS on diverse datasets and collaborated with various domain experts both in the context of (T)BSS (e.g., genome biology, cancer research) as well as outside of it (e.g., ecology, neurology), we think they are very well suited to answer our research questions. Although they cannot provide us with deep data-related insights as they are not application domain experts, they are our primary intended users and bring sufficient experience and a broad perspective to our research questions around TBSS analysis and involved tasks. This helps us to keep \toolname{} generic, yet effective, as was our design goal (\Fref{sec:app}).

No participant uses visual-interactive tools regularly. Their self-assessed experience in visualization is \enquote{basic}, as E4 put it: \enquote{I only use what R has to offer, like ggplot and the base graphics (...) scatterplots, time series plots, (...) box plots. I tend to stick with these basic kinds of plots (...)}.

\subsection{Methodology}
\label{sec:evaluation--methodology}

The interviews were conducted and recorded via Zoom with explicit consent by participants.\footnote{As of manuscript submission, the TU Wien has a Pilot Research Ethics Committee. Approaching it for peer review of research with human participants is not required by the TU Wien, and its response is non-binding. Therefore we do not provide an official ethics approval. Nonetheless, we believe we conducted our research adhering to sufficient ethical standards.} Two researchers were involved in each interview, one tasked with moderation and one took notes. Participants used \toolname{} on their own machines and shared their screen during usage. We used Zoom annotations to point out relevant parts of \toolname{} when necessary.

Both sessions were structured the same. We compiled a text explanation with images of \toolname{}, so that participants can familiarize themselves with it beforehand. The tutorial document was sent to participants together with the consent form ahead of the interview. Steps during the interview were as follows:

\begin{enumerate}
    \item (Only in first session.) We conducted a structured interview about their background and experiences with (T)BSS. 
    \item We gave participants a structured introduction to interactions and visualizations in \toolname{}. The dataset used was synthetic and unfamiliar to them. We asked participants to solve small tasks to practice what we explained. We skipped these tasks when we either saw that they understood it, or when we were short on time.
    \item (Optional.) Participants were allowed to further use \toolname{} for some minutes on their own.
    \item We asked participants to conduct an open analysis on the dataset used in \Fref{sec:datasets--exrates}, which most have worked on in the past, and articulate their thoughts and plans (``think aloud''). We pointed out parts of \toolname{} they did not use or consider so far.
    \item We discussed tasks, visualizations, interactions and possible further improvements in an unstructured fashion. Before we finished the session, we encouraged participants to use \toolname{} more without our supervision.
\end{enumerate}

To answer RQ2 we found it sufficient to check whether or not participants can interpret our visualizations, and if visualizations show the necessary data in the right moment to support their tasks. To do so, we analyzed the recorded video and notes after each session. We looked for articulated suggestions, discussions, and situations where users interacted with visualizations. These instances were transcribed and grouped by tasks (\Fref{sec:task-abstraction--tasks}). Feedback and possible issues of participants were noted, deduplicated, and presented to our collaborators. Subsequent discussions then informed changes to the first design, which we confirmed in the second interview.

The interview guide, tutorial documents, datasets, and our transcripts of the interviews are available as supplementary material.



\subsection{Expert Feedback}
\label{sec:evaluation--expert-feedback}

We describe evidence for our research questions in this section.

\subsubsection{RQ1: Advantages and Disadvantages}
Our participants agreed that \toolname{} has clear advantages compared to current tools used and greatly improves the analysis process. E5 even said that \toolname{} is \enquote{an absolute time saver} and \enquote{very useful for applied work}. The majority of them mentioned that it is easier than in RStudio to compare components, matrices, and parameters. The same outcomes can be achieved faster in \toolname{} and it provides useful new visualizations they could not have in RStudio, such as the component overview/comparison. All participants mentioned to enjoy playing with \toolname{}. Even in our limited time we saw indications how \toolname{} can change the way they work. We observed E4 in the open session to pursue an analysis process resembling binary search, toggling individual lags on and off. Asked about it later, E4 mentioned to be \enquote{not sure if I'd have thought about [this approach] just with RStudio}. E5 was very eager to get hold of \toolname{}, as the intention was to recommend it to their students. E1 stated that better supported comparison tasks give more structure to the analysis process, so all this suggests that \toolname{} allows new or more streamlined analyses.


As for disadvantages, there is one very basic: RStudio allows more flexible and specialized computations than \toolname{}. However, this was not explicitly mentioned by participants. Some said it took time to put everything together, but all our participants managed to do so quickly. A few plots were difficult to understand at first, but after explanations it was relatively easy to use for all participants. In addition, we observed some participants having trouble with idioms that are common and popular in the visualization community, such as PCPs and multiple linked views, which could be overcome by visual literacy efforts.

\subsubsection{RQ2: Supported Tasks}

In this section, we discuss how \toolname{} supports analysis tasks (\Fref{sec:task-abstraction}). We provide quotes from participants to let them speak for themselves, but their sentiment is shared by the majority and not an isolated opinion.


\textit{Identify used parameters (I1):} The tabular overview (\Fref{fig:app--focus--parameters}, A) was considered \enquote{really useful} (E2) and participants thought it \enquote{makes a lot of sense} (E4). 

\textit{Identify unmixing matrix (I2):} Participants could easily identify similar matrices and dominant factors of components. Viewing involved time series (\Fref{fig:app--matrix--exrates}) was considered useful.

\textit{Identify cross-moment diagonality (I3):} It is \enquote{something I don't usually have the time and energy to compute} (E5) and \enquote{very interesting} (E1), but also something they do not regularly use for their analysis today.

\textit{Identify components (I4):} Our participants found the added interactivity compared to RStudio very useful.

\textit{Compare success (C1):} They had no trouble with visual encodings, but participants sometimes forgot that failure is an option.

\textit{Compare parameters (C2):} While the interweaved lag histograms were easy to interpret, it took some time for participants to realize that it is a regular histogram with hidden bins (\Fref{fig:app--interweaved-lags}). Similarity projections of parameters (\Fref{fig:app--focus--parameters}, B) were rarely used by our participants. A possible explanation is because histograms show more data and participants worked with only 5–7 parametrizations, they could use their working memory. We believe their benefits would have become apparent with more parametrizations. 


\textit{Compare unmixing matrices (C3):} Some (E3–E5) mentioned that interpreting the MD-Index for other than extreme values is not easy as it depends on the data dimensionality. While both visualizations were used by all, some participants seemed to prefer the MD-Index (E3, E4) over the factors (E2) to compare matrices.

\textit{Compare component sets (C4):} Participants understood the slope graph (\Fref{fig:app--focus--components}) easily and immediately saw its benefits. E3 mentioned that using it is \enquote{easier than looking at a correlation matrix}.
The projection view was in fact used to see how similar ensemble members are. For this purpose participants also appreciated the component overview (\enquote{you can very fast get an idea of how similar different methods are}, E3), although most did not change the initial clustering parameter.

\textit{Compare possible parameters (C5):} After we introduced participants to individual views and interactions, they learned quickly how to use it and found it useful and convenient. They understood how and why to filter visualized lags, but were not sure about the data-driven calendar-based approach, presumably because they currently analyze data detached from any calendar.
Participants appreciated the PCP with its dimensions, even though they sometimes did not know right away how to interpret all of them: For example, E2 asked what the eigenvalue metric means, what the optimal choice is, and if lower or higher is better. Participants were also sometimes irritated by the number of dimensions, as they depend on the outcome of the refined run.

\subsubsection{RQ3: Possible Improvements}

When asked about improvements to \toolname{}, we got responses mainly pertaining to the parameter selection. E4 would prefer if the syntax to directly select lags matched commands available in R. E2–E4 often ended up with an empty selection in the PCP because they expected brushes to be combined with union instead of intersection. They also want to select all filtered lags and remove all selected lags at once. Aside from the lag selection improvements, more DOI functions would be appreciated. We added one measure for periodicity \cite{vlachos2005} following one participant suggestion.
E5 suggested to support loading precomputed results, possibly from other TBSS methods. E2 asked for more legends, explanations, and a stronger guidance degree. E1 suggested the ability to freely reorder components everywhere, and providing alternative color palettes. With E1 we also discussed the option of showing correlations between input data in the \textit{Input Visualization} screen as another sanity check.


\section{Reflection and Discussion}
\label{sec:conclusion}

Reflecting on our findings and lessons learned during our design study with experts in BSS, we claim that \toolname{} supports tasks involved with TBSS analysis (\Fref{sec:task-abstraction}) and encourages usage of TBSS in various application domains. Despite differences in what an application domain considers interesting in latent dimensions (e.g., doctors might search for specific wave patterns, while investors look for sudden and extreme changes), many tasks are the same. We showed this transferability to financial and medical datasets in \Fref{sec:usage-scenarios}. We developed and evaluated \toolname{} with TBSS experts, who are our primary intended users. They worked with many domain experts in the past to apply TBSS in their respective fields. Their practical experience with different use cases for TBSS informed our visualization design (\Fref{sec:app}). Therefore, based on the mostly positive feedback by our interview participants, we expect that \toolname{} can be useful in many application domains.


In line with the design study methodology \cite{sedlmair2012b}, we used well known visualization idioms and data mining algorithms, applied them in a new context and extended them as necessary. As a consequence, individual parts of \toolname{} will be useful to other visualization researchers and designers. For instance, a slope graph usually shows categorical data cases and their change of rank by line slope. We adapted it to time series by encoding similarity in line thickness. In our user studies it was considered an easy-to-understand visualization to visually compare sets of time series. 
The clustering scheme (\Fref{sec:app--compare-comonents}) is useful whenever members of sets should be clustered and set membership must be taken into account. It works with any dissimilarity measure because it is based on k-medoids. Set-typed data is prevalent \cite{alsallakh2016}, so we expect this to be useful to others.



\subsection{Design Process}
\label{sec:conclusion--design-process}

Following the recommendations of the data-users-tasks design triangle \cite{miksch2014} our proposed visualizations are close to what TBSS experts are used to and therefore quite simple. We also did not include more advanced interactions than highlighting, filtering, hovering, or brushing because TBSS experts come from a text-based software where even these do not exist. Looking back, we think this was a good decision, as in our interviews some participants had initially trouble using, e.g., the PCP. 

What was difficult for us visualization researchers during the design is the domain-independence of TBSS. Our goal, therefore, was to make \toolname{} applicable in a wide range of domain-specific contexts, e.g., in medicine or finance. But both size and complexity of the data vary considerably among the domains, as do the definitions of \enquote{interesting} features and the location and role of TBSS in the data processing pipeline \cite{landesberger2017}. Therefore, we opted in the end for simple interactions and generic/extendable approaches, such as the use of DOI functions, to avoid a \enquote{lock-in} to any specific application domain.

\subsection{Limitations and Future Work}
\label{sec:conclusion--limitations}

We discuss some limitations in our paper. 
Most study participants used the financial dataset (\Fref{sec:datasets--exrates}) at some point in the past to test varying TBSS methods. Although participants fit well to our user description (\Fref{sec:task-abstraction--users}), they were not as intimately familiar with the dataset as it is often the case in visualization-related evaluations. Had this been the case, we may have found additional analysis goals and insights. Nevertheless, we maintain that our study methodology and participant selection was sufficient and appropriate to investigate how \toolname{} impacts involved tasks (\Fref{sec:task-abstraction--tasks}). 
Participants used \toolname{} for around 45~minutes in total on their own terms. More time using it may have surfaced more necessary analysis tasks or improvement suggestions.

As part of our future work, we would like to integrate the suggested improvements by our experts, support larger datasets and allow provision of custom DOI functions.

\section{Summary and Conclusion}
\label{sec:conclusion--conclusion}

We presented \toolname{}, a VA solution for TBSS. TBSS is in a way similar to PCA, in that it can be used to analyze suitable datasets from any application domain, such as biomedical analysis, finance, or civil engineering. Unlike PCA, TBSS properly accounts for temporal correlation and requires complex tuning parameters. Because of these parameter settings, TBSS analysis is inherently open-ended and exploratory as there are no known insights to confirm. \toolname{} is based on a task abstraction and visualization design that we developed together in a user-centered design process with TBSS experts. We evaluated the final interactive prototype with five other TBSS experts, who did not participate in the design process, by conducting two interviews. Feedback from these shows that \toolname{} supports the actual workflow and combination of interactive visualizations that facilitate the tasks involved in analyzing TBSS results—this process was previously a laborious back-and-forth for which analysts had to manually program static visualizations and data mining algorithms. \toolname{} also provides guidance to facilitate the analysis of the data at hand and informed parameter selection, which was previously mostly a guessing game.

\section*{Acknowledgements}

We would like to thank our experts for spending so much of their valuable time on discussions and evaluations with us. This work was supported by the Austrian Science Fund (FWF) under grant P31881-N32.

\bibliography{template}

\end{document}